%% file: main.tex
\documentclass[aps,prc,superscriptaddress,twocolumn,floatfix]{revtex4-1}

\usepackage{amsmath}
\usepackage{graphicx}
\usepackage{dcolumn}
\usepackage{bm}
\usepackage{array}
\usepackage{epstopdf}

\AppendGraphicsExtensions{.tif}

\newcolumntype{M}[1]{>{\centering\arraybackslash}m{#1}}
\newcolumntype{N}{@{}m{0pt}@{}}

\def\lsim{\mathrel{\raise.3ex\hbox{$<$\kern-.75em\lower1ex\hbox{$qf$}}}}
\def\gsim{\mathrel{\raise.3ex\hbox{$>$\kern-.75em\lower1ex\hbox{$\sim$}}}}

\usepackage{amsmath}
\usepackage[margin=1in]{geometry}
\usepackage{amsfonts}
\usepackage{amssymb}
\usepackage{graphicx}
\usepackage[latin1]{inputenc}
\usepackage{siunitx}
\usepackage{tikz}
\usepackage{natbib}
\usepackage{hyperref}
\usepackage{nicefrac}

\newcommand{\isotope}[2]{$^{#2}$#1}
\newcommand{\ns}{\mu}
\def\cogent/{CoGeNT}
\def\mcnp/{MCNPX-PoliMi ver.\ 2.0}
\def\NI/{National Instruments}
\DeclareSIUnit{\keVnr}{\text{keV}\ensuremath{_{\text{nr}}}}
\DeclareSIUnit{\keVee}{\text{keV}\ensuremath{_{\text{ee}}}}
\DeclareSIUnit{\curie}{\text{Ci}}
\DeclareSIUnit{\sample}{\text{S}}
\DeclareSIUnit{\mm}{\text{mm}}
\DeclareSIUnit{\kg}{\text{kg}}
\DeclareSIUnit{\cm}{\text{cm}}
\DeclareSIUnit{\keV}{\text{keV}}
\DeclareSIUnit{\us}{$\mu$s}
\DeclareSIUnit{\MeV}{\text{MeV}}

\begin{document}

\preprint{Preprint XYZ}

\title{Measurement of the low-energy quenching factor in germanium\\ using an $^{88}$Y/Be photoneutron source}

\author{B.J. Scholz}
\thanks{Corresponding author: scholz@uchicago.edu}
\affiliation{Enrico Fermi Institute, Kavli Institute for Cosmological Physics, and Department of Physics, University of Chicago, Chicago, IL 60637, USA}

\author{A.E. Chavarria}
\affiliation{Enrico Fermi Institute, Kavli Institute for Cosmological Physics, and Department of Physics, University of Chicago, Chicago, IL 60637, USA}

\author{J.I. Collar}
\affiliation{Enrico Fermi Institute, Kavli Institute for Cosmological Physics, and Department of Physics, University of Chicago, Chicago, IL 60637, USA}

\author{P. Privitera}
\affiliation{Enrico Fermi Institute, Kavli Institute for Cosmological Physics, and Department of Physics, University of Chicago, Chicago, IL 60637, USA}

\author{A.~E.~Robinson}
\thanks{Present address: Fermi National Accelerator Laboratory, Batavia, Illinois 60510, USA}
\affiliation{Enrico Fermi Institute, Kavli Institute for Cosmological Physics, and Department of Physics, University of Chicago, Chicago, IL 60637, USA}

\date{\today}

\begin{abstract}
We employ an $^{88}$Y/Be photoneutron source to derive the quenching factor for neutron-induced nuclear recoils in germanium, probing recoil energies from a few hundred eV$_\text{nr}$ to \SI{8.5}{\keVnr}. A comprehensive Monte Carlo simulation of our setup is compared to experimental data employing a Lindhard model with a free electronic energy loss $k$ and an adiabatic correction
for sub-keV$_\text{nr}$ nuclear recoils. The best fit $k=0.179\pm 0.001$ obtained using a Monte Carlo Markov Chain (MCMC) ensemble sampler is in good agreement with previous measurements, confirming the adequacy of the Lindhard model to describe the stopping of few-keV ions in germanium crystals at a temperature of $\sim$77 K. This value of $k$ corresponds to a quenching factor of \SI{13.7}{\percent} to \SI{25.3}{\percent} for nuclear recoil energies between \SI{0.3}{\keVnr} and \SI{8.5}{\keVnr}, respectively.

\end{abstract}

\pacs{Valid PACS appear here}
\maketitle

\section{Introduction}

Weakly Interacting Massive Particles (WIMPs), hypothetical particles able to account for most observations pointing at a cosmological dark matter, are expected to interact via elastic scattering off nuclei in detecting media. Detector signals would arise from the energy loss of the recoiling nucleus as it slows down. The interpretation of WIMP searches crucially depends on a correct understanding of the mechanisms governing the stopping of low-energy ions in the target material. This concern can be extended to experimental efforts aiming to measure coherent elastic neutrino-nucleus scattering \cite{coherent}, where the mode of interaction and energy regime are the same. 

At the few-keV energies expected from WIMP or low-energy neutrino interactions, nuclear recoils typically induce a smaller response  than electron recoils of the same energy. Depending on detector type, this response is often measured through the scintillation or ionization yield. In the case of standard germanium diodes operated at liquid nitrogen temperature, it is the second mechanism that is exploited to extract signals. An energy-dependent quenching factor can then be defined as the ratio between the ionization generated by the recoil of a germanium nucleus, and that from an electron recoil of the same energy. 

We report on a new measurement of the germanium quenching factor at $\sim$77 K, using a P-type Point Contact (PPC) detector \cite{barbeau-01}, and a calibration technique recently described in \cite{collar-01}. This approach employs a photoneutron radioactive source, exploiting its monochromatic low-energy neutron emission to create nuclear recoils having a well-defined maximum recoil energy of just a  few keV$_\text{nr}$ (the suffix stands for ``nuclear recoil", as opposed to the  smaller ``electron equivalent" (ee) ionization energy that is actually measured post-quenching). The modest electronic noise characteristic of a PPC allows to include the contribution from sub-keV$_\text{nr}$ nuclear recoils. This technique has been used thus far in the characterization of the quenching factor of sodium recoils in NaI(Tl) scintillators \cite{collar-01}, and carbon and fluorine recoils in superheated fluids \cite{piconim,alanthesis,pico60}. 

\section{Experimental Setup}

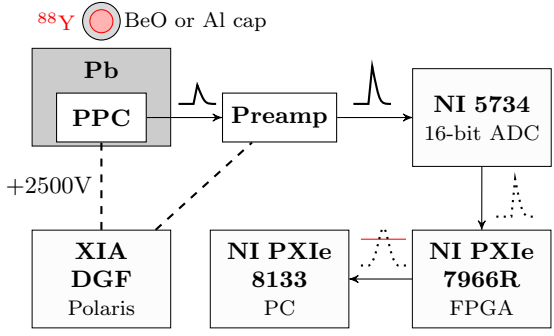
\begin{figure}[tbp]
\input{./Plots/daq.tex}
\caption{Experimental setup: the preamplifier output is digitized using a NI 5734 16-bit ADC, and shaped with a digital trapezoidal pulse shaper implemented on a NI 7966R Field-Programmable Gate Array (FPGA). The preamplifier trace is stored on the host PC if the corresponding shaped signal triggers on a rising edge threshold set at $\sim$\SI{0.8}{\keVee}, also implemented in the FPGA.}
\label{fig:daq}
\end{figure}

 The experimental arrangement is illustrated in Fig. \ref{fig:daq}. All measurements took place in a shallow underground laboratory (6 m.w.e.) at the University of Chicago. A $\SI{50.7}{\mm}$ (diameter) $\times~ \SI{43.0}{\mm}$ (length) PPC germanium detector manufactured by Canberra Industries with an original active mass of \SI{0.475}{\kg} was surrounded by \SI{20}{\cm} of lead. This shielding reduces the intense gamma emissions from the source to a manageable level, avoiding pile-up and data throughput limitations, while causing only minimal changes to neutron energies \cite{collar-01}. The detector was previously used by the \cogent/ collaboration \cite{aalseth-01, aalseth-02}. An \isotope{Y}{88} gamma source was encapsulated by a $\SI{1}{cm}$-thick gamma-to-neutron BeO converter, and placed \SI{23}{\cm} away from the front of the PPC detector.  The dominant neutron energy emitted by the source is $E_{n}\,=\,$\SI{152}{\keV} with an additional  small ($\SI{0.5}{\percent}$) component of $E_{n}\,=\,\SI{963}{\keV}$ \cite{collar-01}. The maximum nuclear recoil energy transferred within a single scatter event in Ge for these neutron energies is $E_\text{nr}^{max}=(4MmE_{n})/(M+m)^2=\SI{8.5}{\keVnr}$ and $E_\text{nr}^{max}=\SI{51}{\keVnr}$, respectively, where $M$ and $m$ stand for Ge nucleus and neutron masses. 
 
 A \isotope{He}{3} neutron counter surrounded by HDPE moderator was employed to measure the isotropic neutron yield of the source, found to be in the range 574-580 neutrons/s, depending on the orientation of the source with respect to the counter. Prior experience with this \isotope{He}{3}  counter and other neutron sources (\isotope{Am}{241}/Be, \isotope{Pu}{239}/Be, \isotope{Cf}{252}) of known activity point at an ability to characterize their yield within a few percent of its nominal value. More specifically, seven previous measurements involving four different commercial neutron sources displayed a systematic trend to underestimate their nominal neutron yield by $\sim$12\% \cite{drew}. The activity of the source was separately assessed via a gamma emission measurement employing a dedicated coaxial germanium detector. This  gamma yield was used as an input to a  \mcnp/ \cite{mcnpx} simulation employing a revised cross-section \cite{alan2} for the $^{9}$Be($\gamma$,n)$^{8}$Be reaction. The neutron yield obtained via this simulation is compatible with \isotope{He}{3} counter measurements, at  $\sim573$ neutrons/s. Combining all measurements and accounting for statistical, simulation, and cross-section uncertainties, we estimate a source activity of 0.640$\pm$4\% mCi, corresponding to an emission of 574$\pm$5\% neutrons/s.
 
Preamplifier power and detector high voltage to the PPC were provided by a Polaris XIA DGF. The preamplifier signal output was fed into a 16-bit National Instruments (NI) 5734 ADC, connected to an NI PXIe-7966R FPGA module. The host PC was a NI PXIe 8133. A trapezoidal, digital pulse shaper was implemented on the FPGA using the recursive algorithm in \cite{jordanov-01}. The total shaping time was set to \SI{16}{\us} with a peaking time of $\SI{8}{\us}$ and a zero length flat top. A rising edge threshold trigger set to approximately \SI{0.8}{\keVee} was used for real-time detection of digitally-shaped pulses. The trigger position was set to \SI{80}{\percent} of the \SI{400}{\us}-long waveforms, with a sampling rate set to \SI{40}{\mega\sample\per\second}. The \SI{320}{\us}-long pre-trigger trace allowed monitoring of detector noise and baseline stability. An electron-equivalent energy scale was established using the \SI{59.5}{\keV} $\gamma$-emission from \isotope{Am}{241}, as well as the four main emission lines from \isotope{Ba}{133}.

\begin{figure}[tbp]
\includegraphics[width=\linewidth]{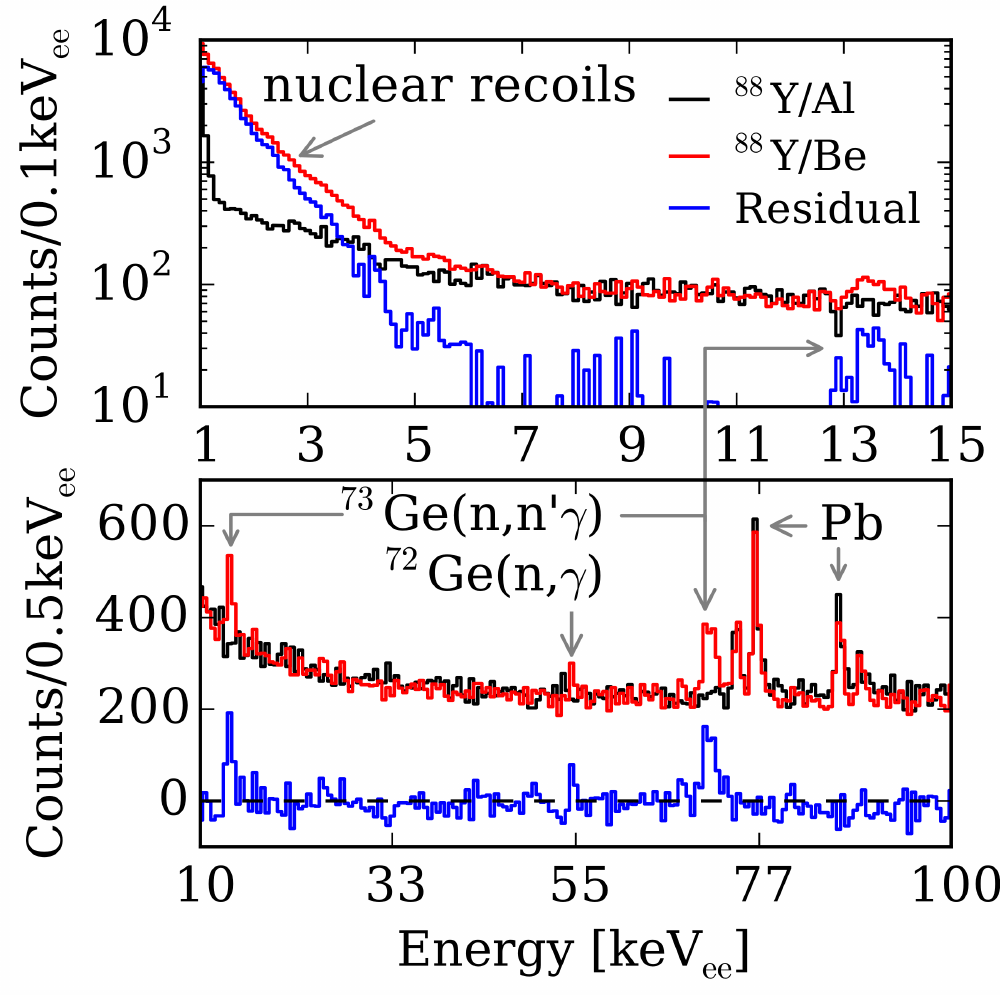}
\caption{Normalized energy spectra recorded for the two different source configurations. Their difference (residual) is shown in blue. The digitizer gain setting limited usable data to $>1$ keV$_{ee}$. The low-energy residual excess arises from neutron-induced nuclear recoils. Additional neutron-induced signals are visible at \SI{13.3}{\keV},  \SI{53.3}{\keV}, and \SI{68.8}{\keV}. These peaks are the result of $^{72}$Ge$(n,\gamma)$ and $^{73}$Ge$(n,n^\prime\gamma)$ interactions \cite{jones-02,clover}. The cancelation of Pb fluorescence lines in the range \SI{72}{\keV_{ee}} to \SI{87}{\keV_{ee}} illustrates the absence of isolated x/$\gamma$-ray contributions to the residual spectrum.}
\label{fig:spectra}
\end{figure}

In order to separate neutron-induced signals from those generated by gamma interactions from the source, a second measurement was performed where the BeO converter was replaced by an aluminum cap of identical geometry. Aluminum has a total attenuation for dominant (898 keV) \isotope{Y}{88}  gamma-rays  of $\lambda_\text{Al}(\SI{1}{\MeV}) = \SI{0.06146}{\cm\squared\per\gram}$, which closely matches that from BeO, $\lambda_\text{BeO}(\SI{1}{\MeV}) = \SI{0.06112}{\cm\squared\per\gram}$ \cite{berger-01}. A total \SI{19.3}{\hour} of exposure with the \isotope{Y}{88}/BeO source configuration and  \SI{20.0}{\hour} with \isotope{Y}{88}/Al were collected. The energy spectra are normalized to account for the difference in run times, and the decay of the source (T$_{\nicefrac{1}{2}}$ = \SI{106.65}{\day}). The residual spectrum, i.e. the difference between the \isotope{Y}{88}/BeO (gammas and neutrons) and \isotope{Y}{88}/Al (gammas) spectra contains neutron-induced signals only \cite{collar-01}. Fig. \ref{fig:spectra} shows both normalized spectra, and the resulting residual spectrum. The low-energy excess in the residual is caused by neutron-induced germanium recoils. As expected, the residual rapidly converges to zero above few keV$_{ee}$, except for discrete peaks arising from inelastic scattering and neutron capture in $^{72,73}$Ge \cite{jones-02,clover}. These peaks can display a characteristic asymmetry towards high energies, due to the addition of gamma and nuclear recoil energy depositions \cite{skoro,jova}.

\begin{figure}[tbp]
\includegraphics[width=\linewidth]{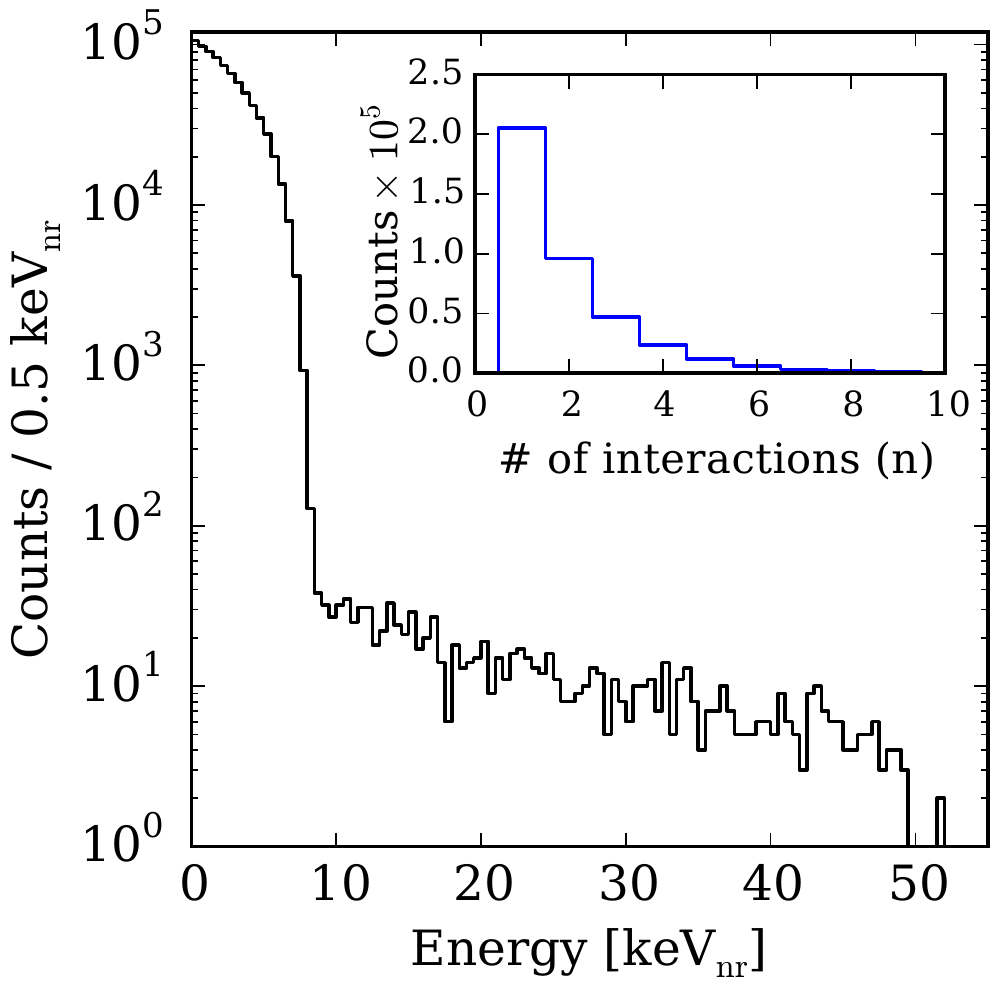}
\caption{Simulated, unquenched distribution of nuclear recoil energies deposited for each individual neutron scatter event. As expected, primary $E_{n}\,=\,$\SI{152}{\keV} neutrons produce recoil energies of up to \SI{8.5}{\keV_{nr}}. The 0.5\% $E_{n}\,=\,\SI{963}{\keV}$ branch contributes a small fraction of higher recoil energies up to \SI{51}{\keV_{nr}}. The inset shows the multiplicity of interactions in the PPC for all simulated neutron histories. }
\label{fig:unquenched-recoil-spectrum}
\end{figure}

In addition to these measurements, a total of $10^8$ neutrons emitted by the BeO converter was simulated using \mcnp/ \cite{mcnpx}. The geometry included fine details such as the known internal structure of the PPC, chemical impurity content of lead, and source encapsulation. It also involved new improved cross-section libraries specifically developed for dark matter detector simulations \cite{alan}. Approximately \SI{0.4}{\percent} of these simulated neutrons produce at least one recoil within the detector. The interaction depth, measured from the nearest surface of the germanium crystal, and recoil energy from each nuclear elastic scattering event were recorded. The unquenched energy distribution of these individual recoils is shown in Fig.
\ref{fig:unquenched-recoil-spectrum}.  Approximately \SI{50}{\percent} of neutrons interacting with the germanium crystal do so only once, a fraction large enough to expect a readily visible endpoint energy in the ionization spectrum, corresponding to the expected maximum recoil energy transfer of  \SI{8.5}{\keV_{nr}}. Multi-scatter events allow to study the contribution from nuclear recoils individually depositing energies below the 0.8 keV$_\text{ee}$ triggering threshold (Fig. \ref{fig:ss-ms-vs-res}). More precisely, 30(15)\% of simulated neutrons interacting with the detector produce at least one recoil depositing less than 1(0.5) keV$_\text{nr}$.

\section{Analysis}
To extract the quenching factor we compare the simulated data to the experimental residual spectrum. In a first step, the energy deposition of each simulated nuclear recoil is converted into an electron-equivalent energy via an energy-dependent quenching model $Q(E_\text{nr})$. Previous measurements of the quenching factor in germanium suggest that the Lindhard theory \cite{lindhard-01} provides an adequate description of $Q$ down to very low energies. This formalism can be written as \cite{barker-01,benoit-01}
\begin{align}
Q&= \frac{k\,g(\epsilon)}{1+k\,g(\epsilon)}\label{eq:lindhard-model-1}\\
g(\epsilon) &= 3\,\epsilon^{0.15} + 0.7\,\epsilon^{0.6}+\epsilon\label{eq:lindhard-model-2}\\
\epsilon &= 11.5\,Z^{-\nicefrac{7}{3}}\,E_\text{nr}\label{eq:lindhard-model-3}.
\end{align}
Here $Z$ is the atomic number of the recoiling nucleus, $\epsilon$ a dimensionless energy, $E_\text{nr}$ is the recoil energy in keV$_\text{nr}$, and $k$ describes the electronic energy loss. In the original description by Lindhard, a value $k\,=\,0.133 Z^{\nicefrac{2}{3}} A^{-\nicefrac{1}{3}}\,(=\,0.157$ for Ge) was adopted, with $A$ the mass number of the nucleus. Lindhard-like models have been fitted to previous quenching factor measurements using comparable  $k$ values \cite{barker-01,hooper-01}. Accordingly, we treat $k$ as the free parameter of prime interest in our analysis. 

In a second step, we acknowledge that the charge collection efficiency $\eta$ within a PPC detector varies with interaction depth into the crystal. This is due to the effect of a lithium-diffused external contact covering most of the outer surface of the diode \cite{aalseth-04}. Following \cite{aalseth-03} we adopt a sigmoid-shaped charge collection efficiency profile
\begin{align}
\eta(x,\delta,\tau)\;=\;1-\frac{1}{\exp\left[{\frac{x-(\delta+0.5\,\tau)}{0.17\,\tau}}\right]+1},
\end{align}
where $\delta$ is an outermost dead layer thicknesses for which $\eta$ is negligible. $\tau$ is an underlaying transition layer thickness over which the charge collection efficiency rises from $\eta=0.05$ to $0.95$, and $x$ is the interaction depth. 

In a third step, we account for the possibility of a reduced ionization
efficiency for slow-moving nuclear recoils, by introducing a smooth
adiabatic correction factor $F_\text{AC}$ to the Lindhard stopping. The concept of a 
"kinematic threshold"  below which 
the minimum excitation energy of the detector system is larger than
the maximum possible energy transfer to an electron by a
slow-moving ion, can be traced back to Fermi and Teller \cite{fermi}.
We adopt the same correction factor model previously employed in
\cite{ahlen-01,ahlen-02},
\begin{align}
F_\text{AC}\left(E_\text{nr},\xi\right) = 1 -
\text{exp}\left[-\nicefrac{E_\text{nr}}{\xi}\right],
\end{align}
where the adiabatic energy scale factor $\xi$ corresponds to the threshold energy below which a rapid drop in ionization efficiency can be expected. 


The total simulated electron equivalent energy measured for a neutron interacting $n$ times with the crystal can now be written as 
\begin{align}
E_\text{ee} = \sum\limits_{i=1}^n
E_\text{nr}^{(i)}Q\left(E_\text{nr}^{(i)},k\right)\eta\left(x^{(i)},\delta,\tau\right)F_\text{AC}\left(E_\text{nr}^{(i)},\xi\right)\label{eq:electron-equivalent-energy},
\end{align}
where  $E_\text{nr}^{(i)}$ is the recoil energy deposited at the $i^{th}$ interaction site. The resulting nuclear recoil energy spectrum in units of electron equivalent energy is convolved with a resolution $\sigma^2(E_\text{ee}) = (\SI{69.7}{eV})^2 + 0.98$ eV $E_\text{ee}$(eV), specific for this detector \cite{aalseth-01,aalseth-02}. 

In a final step, the simulated spectrum is normalized to match the integrated neutron yield over the time span of the measurements. To account for the mentioned significant uncertainty in source neutron yield we introduce an additional free global scaling parameter $\gamma$. Our full analysis therefore involves a total of five free parameters, three of which ($\delta,\tau,\gamma$) are treated as nuisance parameters as they are not of immediate interest to our measurement of the quenching factor, even if they must be accounted for.

We employ a Monte Carlo Markov Chain (MCMC) to find the parameter set $\vec{\pi} = \left(k,\delta,\tau,\xi,\gamma\right)$ that provides the best fit of the simulated data to the experimental residual spectrum. Assuming an underlying Poisson distribution for each bin of the simulated residual spectrum, the probability to count $N_i$ events in bin $i$ given $\ns_i$ simulated counts in the same bin can simply be written as
\begin{align}
P(N_i|\ns_i)\;=\;\frac{\ns_i^{N_i}\,\text{e}^{-\ns_i}}{N_i!},
\end{align}
where $\ns_i$ solely depends on our choice of fit parameters $\vec{\pi}$. The corresponding log-likelihood function is given by
\begin{align}
\ln \text{L}(\vec{N}|\vec{\pi}) = & \sum\limits_i N_i\ln(\ns_i(\vec{\pi})) - \sum\limits_i \ns_i(\vec{\pi})\label{eq:log-likelihood}\\
& - \sum\limits_i \ln(N_i!).\nonumber
\end{align}
The last sum is constant for all choices of $\vec{\pi}$. We will therefore not include it in the final posterior probability sampling process. From Bayes' theorem we know that 
\begin{align}
P(\vec{\pi}|\vec{N}) \propto P(\vec{N}|\vec{\pi})P(\vec{\pi}), \label{eq:bayes-theorem}
\end{align}
with
\begin{align}
P(\vec{\pi}) = P(k)P(\delta)P(\tau)P(\xi)P(\gamma), \label{eq:independent-parameters}
\end{align}
where we assume that all parameters are independent. For our analysis we choose a bound, flat prior for each parameter (Table  \ref{tab:fit-parameters}) for their respective limits. Neglecting the normalization constant of Eq. (\ref{eq:bayes-theorem}), the final logarithmic posterior probability distribution can be written as
\begin{align}
\ln P(k,\delta,\tau,\xi,\gamma|\vec{N}) = &\ln L(\vec{N}|k,\delta,\tau,\xi,\gamma)\label{eq:posterior}\\
& + \ln P(k,\delta,\tau,\xi,\gamma).\nonumber
\end{align}
The last logarithm is either 0 or $-\infty$, depending on whether all parameters are within their respective bounds or not. We use \textbf{emcee} \cite{goodman-02}, a pure Python implementation of Goodman and Weare's affine invariance ensemble sampler \cite{goodman-01} to sample Eq. (\ref{eq:posterior}).

\begin{figure}[tbp]
\includegraphics[width=\linewidth]{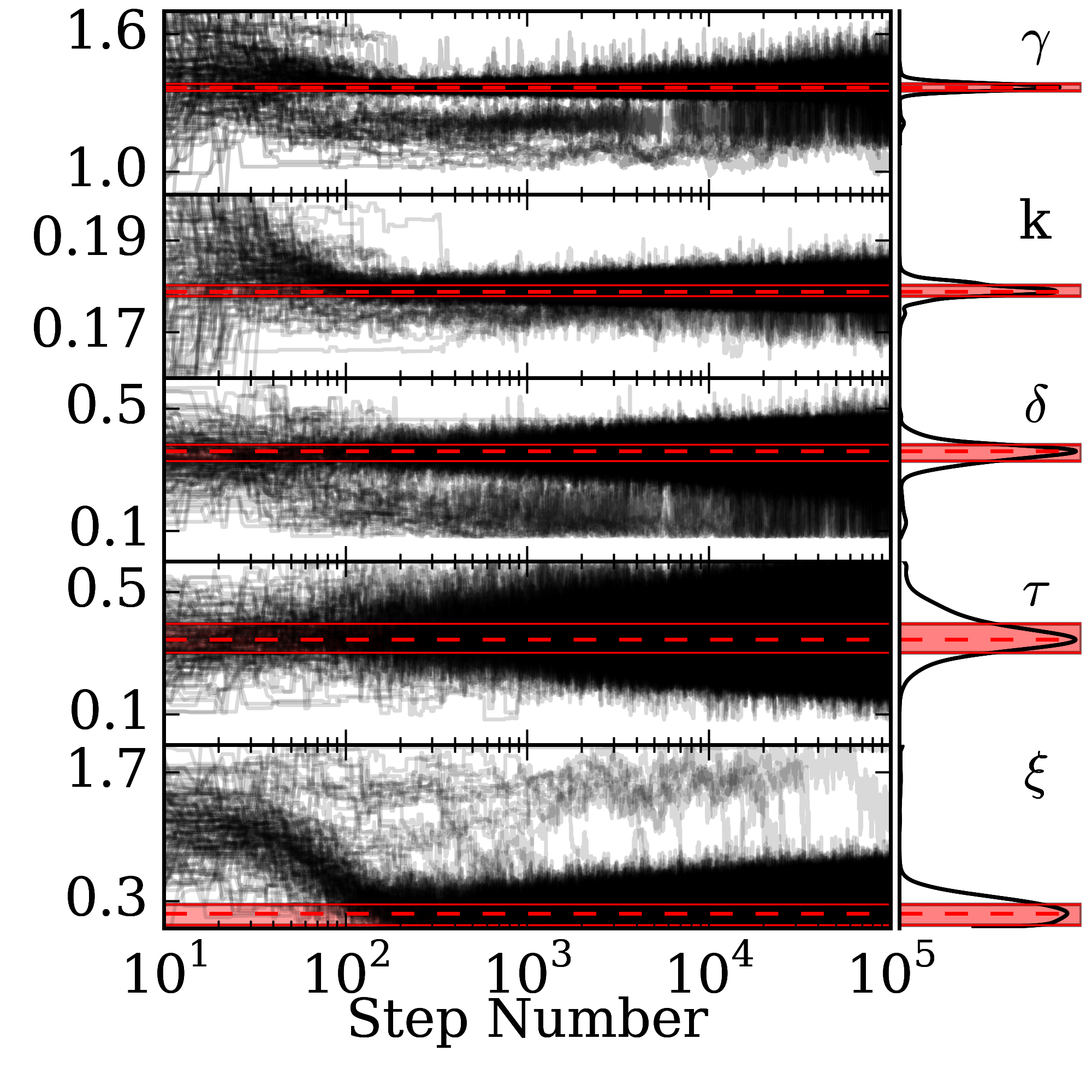}
\caption{Full MCMC chain consisting of 320 walkers with $10^5$
iterations each. The walkers were initialized uniformly within the
allowed parameter limits (Table \ref{tab:fit-parameters}). Most
walkers converge onto their final probability distribution after
$\sim500$ steps. The right-side plots show the kernel density
estimation using a bandwidth chosen according to Silverman's rule
\cite{silverman-01}. The dashed red line highlights the most probable
value of the resulting marginalized posterior probability
distribution function. The shaded red area shows the $1\sigma$
credible region.}
\label{fig:mcmc-parameter-distribution}
\end{figure}

\section{Results}

The first MCMC run performed consists of 320 walkers with $10^5$
steps each. The walkers are initialized uniformly within the allowed
parameter space. The full chain is shown in Fig.\
\ref{fig:mcmc-parameter-distribution}. Most walkers are observed to
converge onto the target distribution after $\sim500$ steps. The
adiabatic energy scale factor $\xi$  exhibits the longest auto-correlation time
with $\tau_\text{acor}\approx 87$ steps. The full chain therefore
covers a total of approximately 1150 auto-correlation lengths,
whereas the burn-in time is limited to the first six. Following
\cite{sokal-01} we choose to discard the first twenty
$\tau_\text{acor}$ to eliminate any remaining initialization bias.
The mean acceptance probability for the remaining chain is
P$_\text{acc}=0.46$. All parameters show a monotonically decreasing
Gelman-Rubin potential scale reduction factor R$_\text{GR}$
\cite{gelman-01}, the largest of which is R$_\text{GR}(\xi)$ = 1.073
after $10^5$ steps. An additional visual inspection of all walker
trajectories suggests proper mixing within each chain.  The marginalized best-fit values including
their $1\sigma$ credible region are provided in Table
\ref{tab:fit-parameters}. To further investigate the presence of any 
possible meta-stable states, we run three additional, shorter MCMC
chains of 320 walkers and $2\times10^4$ steps with differing starting
conditions. For the first two additional runs all
parameters are set below, or above, their respective best-fit values (Table \ref{tab:fit-parameters}). The third run probes
a possibly meta-stable state visible at $\xi\approx\SI{1.65}{\keVnr}$ in Fig.\
\ref{fig:mcmc-parameter-distribution} by initializing all walkers within the
vicinity of $\xi=\SI{1.65}{\keVnr}$, whereas all other parameters are
uniformly distributed within their respective bounds. All three runs
converge onto the same posterior distribution as the initial MCMC
run. The burn-in times, mean acceptance fractions and
auto-correlation lengths are generally identical. We conclude that the investigated possibly meta-stable state
bears no significance, and that all walkers have properly explored the
phase space and fully stabilized on the final posterior probability
distribution.

\renewcommand{\arraystretch}{1.5}
\begin{table}[tbp]
\begin{tabular}{lcc}
\\
\hline
\hline
Parameter & Boundaries & Best Fit\\
\hline
$k$ & $\left[0.1,0.3\right]$ &$0.1789^{+0.0014}_{-0.0010}$\\[5pt]
$\delta$ [mm] & $\left[0.5,6.0\right]$
&$3.60^{+0.22}_{-0.31}\,$\\[5pt]
$\tau$ [mm] & $\left[0.5,6.0\right]$ & $3.44^{+0.53}_{-0.43}\,$\\[5pt]
$\xi$ [keV$_{nr}$] & $\left[0.0,2.0\right]$ &
$0.16^{+0.10}_{-0.13}\,$\\[5pt]
$\gamma$ & $\left[0.5,2.5\right]$ & $1.367^{+0.015}_{-0.014}$\\
\hline
\hline
\end{tabular}
\caption{Parameter space and marginalized best-fit values for all
free parameters. The errors provided represent the $1\sigma$ credible
region obtained from the MCMC analysis. The upper boundary on the
explored adiabatic energy scale ($\xi$) space has been chosen
arbitrarily, but large enough such that it does not affect walker
movement.}
\label{tab:fit-parameters}
\end{table}

\begin{figure}[tbp]
\includegraphics[width=\linewidth]{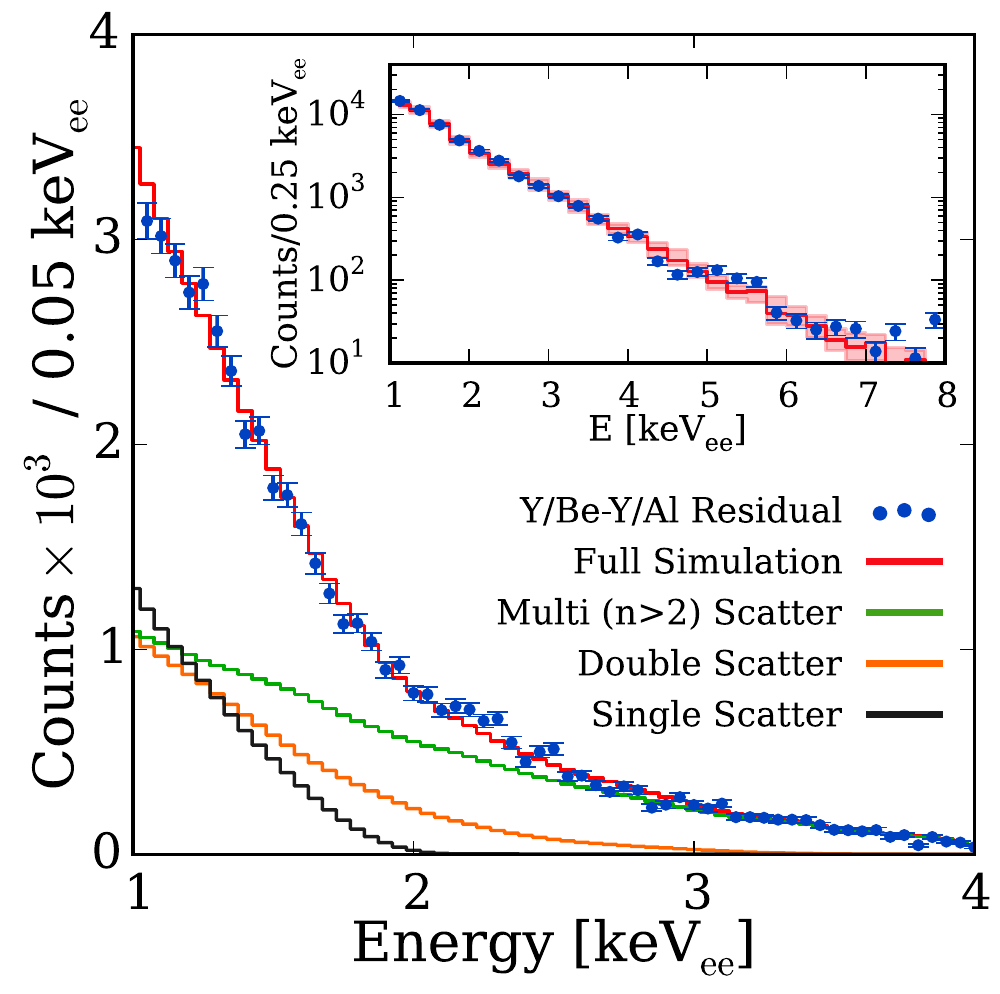}
\caption{Contributions from single and multiple neutron scattering interactions to the measured ionization energy spectrum. Below \SI{2}{\keV_{ee}}  single, double, and multi-scatter  ($n>$2) events contribute approximately the same to the overall spectrum. The endpoint of the single scatter spectrum corresponds to an energy of approximately \SI{2}{\keVee}, as expected from previous measurements of the germanium quenching factor at 77 K. This endpoint is readily visible as an inflection in the experimental residual. The shaded red band in the inset shows the one-sigma credible band for the fit. The quality of the fit is $\chi^2/\text{d.o.f}$ = \nicefrac{19.3}{13}. }
\label{fig:ss-ms-vs-res}
\end{figure}

The most probable value of $k=0.1789$ is close to the semi-empirical prediction by Lindhard of $k=0.157$, previous modeling and fits \cite{barker-01,hooper-01}, and in good agreement with existing experimental data at discrete energies. Below \SI{0.8}{\keVnr} our
quenching model starts to deviate from a pure Lindhard model due to
the adiabatic correction factor $F_\text{AC}$. The
corresponding best-fit value of the adiabatic energy scale factor
$\xi=\SI{0.16}{\keVnr}$ is seen to be in good agreement with  kinematic threshold predictions recently made for germanium
\cite{sorensen-01}. As discussed above, $\xi$ lies well below our triggering threshold of $\sim\SI{0.8}{\keVee}$. However, our simulations show that approximately one third of the triggering events between 1-\SI{2}{\keVee} involve three or more interactions with the detector (Fig. \ref{fig:ss-ms-vs-res}). The cumulative ionization energy from events involving multiple scatters can surpass the triggering threshold, contributing to the experimental residual. The energy range for which
our analysis provides a valid description of the quenching factor is
limited from above by the maximum recoil energy from a single
(dominant branch) neutron scatter,
$E_\text{nr}^\text{max}=\SI{8.52}{\keVnr}(\approx\SI{2.15}{\keVee})$.\\

\begin{figure}[tbp]
\includegraphics[width=\linewidth]{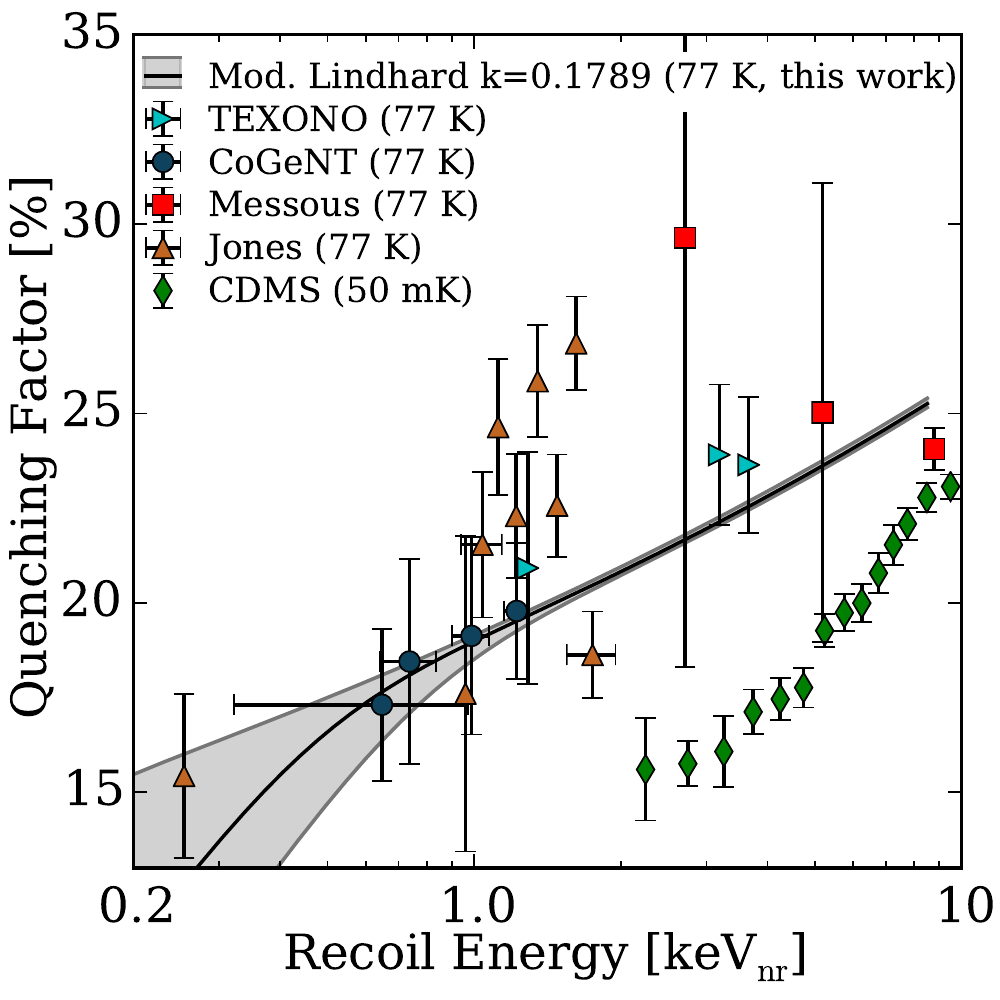}
\caption{Best-fit germanium quenching factor obtained from this work. Data points correspond to previous measurements from \cite{barbeau-01,chasman-01,chasman-02,jones-01,messous-01,texono} in this recoil energy region at 77 K. The solid line shows the modified
Lindhard model for our best-fit $k=0.1789$ and $\xi=\SI{0.16}{\keVnr}$,
over the energy region probed by this calibration. Below
$\sim$\SI{0.8}{\keVnr} the quenching factor is affected by the adiabatic correction factor $F_\text{AC}$. The
maximum recoil energy probed is given by the maximum energy transfer
of a single (dominant branch) neutron scatter, i.e.
$\SI{8.5}{\keVnr}$. Grayed lines represent the combined $1\sigma$
credible region for $k$ and $\xi$. Additional data points at 50 mK are shown \cite{CDMS}. See text for a discussion on a possible temperature dependence for this quenching factor.}
\label{fig:equivalent-energies}
\end{figure}

The best-fit overall scaling $\gamma=1.367$ would suggest a neutron yield
from the source \SI{36.7}{\percent} larger than measured with the
\isotope{He}{3} counter. This best-fit value was found
to be robust ($\pm^{2.3}_{2.1}\%$) against small ($\pm 7\%$)
variations in the magnitude of the neutron cross-section in lead,
representative of its known uncertainty. We performed a similar study of the dependence of $\gamma$ on the $\pm 5\%$ estimated uncertainty in germanium cross-sections, and $\pm 20\%$ uncertainty in the strength function (a measure of resonance contribution) for this element.  These result in an additional variation in $\gamma$ by $\pm^{3.8}_{5.0}\%$. The obtained best-fit value for $\gamma$ is deemed satisfactory, in
view of the uncertainties involved, and in particular the mentioned tendency for our \isotope{He}{3} measurements to underestimate the nominal neutron yield from commercial sources.  In addition to this, an anti-correlation between the active volume of the detector (i.e., the bulk unaffected by dead or transition layer) and $\gamma$ exists.  This active volume changes rapidly with the adopted value of $\delta$ and $\tau$, e.g., already by $\sim$15\% over the uncertainty in their best-fit values (Table  \ref{tab:fit-parameters}). While this correlation is unavoidable, the best-fit values of $\delta$ and $\tau$ can be contrasted with expectations, as follows. The thickness of these layers was measured soon after detector acquisition in 2005, using an uncollimated \isotope{Am}{241} source, finding them similar at $\sim$1.2 mm each \cite{aalseth-03}. This was in line with the deep lithium diffusion requested from the manufacturer. Lithium diffusion in the external n+ contact in P-type germanium detectors is known to progress in time, specially for crystals stored at room temperature, as has been the case for most of this detector's history. Based on the few available measurements for this evolution (an increase in thickness by factors 3.3 (4.2) over 9 (13) years \cite{huy-01,huy-02}) we allowed a large parameter space for $\delta,\tau \in \left[0.5\,\text{mm},6\,\text{mm}\right]$. The obtained best-fit values for $\delta$ and $\tau$ correspond to an increase in the sum of dead and transition layer thicknesses in our PPC by a factor of 2.9 over a decade, compatible with the observations in \cite{huy-01,huy-02}. 

The quenching factor corresponding to our best-fit $k=0.1789$ and
$\xi=\SI{0.16}{\keVnr}$ is shown in Fig. \ref{fig:equivalent-energies}. A good agreement with previous measurements at 77 K  is evident.  Fig. \ref{fig:ss-ms-vs-res} shows a comparison of best-fit simulated recoil spectrum and experimental residual over the 1-8 keV$_{ee}$ fitting range.

\section{Conclusions}

We have demonstrated a new calibration method described in \cite{collar-01}, expanding its use to germanium targets at 77 K, finding  an excellent agreement with previous quenching factor measurements at discrete recoil energies. The simplicity of the experimental setup, combined with a straightforward data analysis, invites to apply this method to other WIMP and neutrino detector technologies. The emitted neutron energy can be adjusted by replacing the \isotope{Y}{88} source with other suitable isotopes such as \isotope{Sb}{124} ($E_n=\SI{24}{\keV}$) or \isotope{Ba}{207} ($E_n=\SI{94}{\keV}$). In upcoming publications we will report on results already obtained for silicon recoils in CCDs \cite{alvaro}, and xenon recoils in a single-phase liquid xenon detector \cite{luca}. 

Recent work \cite{dm,rom} points at a possible dependence of the low-energy quenching factor in germanium on detector temperature and internal electric field, potentially related to the disagreement between all present results at 77 K, and those obtained at 50 mK \cite{benoit-01,CDMS,shutt} (Fig. \ref{fig:equivalent-energies}). This disagreement must be understood, as it might impact the physics reach of competing detector technologies. Use of the presently described technique on cryogenic germanium detectors \cite{cdmstalk} should help clarify the origin and extent of these discrepancies.

\begin{acknowledgements}
This work was supported in part by the Kavli Institute for Cosmological Physics at the University of Chicago through grant NSF PHY-1125897 and an endowment from the Kavli Foundation and its founder Fred Kavli. It was also completed in part with resources provided by the University of Chicago Research Computing Center.
\end{acknowledgements}

\bibliographystyle{plain}
\bibliography{Ge-Quenching-Bib}

\end{document}

%% file: Plots/daq.tex
\tikzstyle{rect} = [rectangle, draw, text centered, rounded corners]
\tikzstyle{block} = [rectangle, very thin, draw,fill=black!1, text width=5em, text centered, minimum height=4em]
\tikzstyle{cloud} = [draw, red, ellipse,fill=black!10, node distance=3cm, minimum height=2em]
\tikzstyle{line} = [draw, -stealth']
\usetikzlibrary{shapes,arrows,positioning}

\begin{tikzpicture}[node distance = 1cm, auto]
	\node [circle, thin, draw, black, fill=black!15, minimum width = 0.5cm] (converter) {};
	\node [circle, thin, draw,red, fill=red!30, minimum width = 0.25cm] (source) {};
    \node [red] at (-0.6,0) {\footnotesize{\isotope{Y}{88}}};   
    \node [black] at (1.25,0) {\footnotesize{BeO or Al cap}};   
	\node [rectangle,fill=black!20,very thin, draw, text width=5em, text centered, below=0.1cm of converter,minimum height=4em] (lead) {\textbf{Pb}\vspace{2.25em}};
	\node [rectangle,draw,below=-2.2em of lead,text width=3em,fill=white,text centered,minimum height=2em] (detector) {\textbf{PPC}};
	\node [rectangle,draw,text width=4em,text centered,minimum height=2em,right=of detector] (preamp) {\textbf{Preamp}};
	\node [block,right=of preamp] (NI5734) {\textbf{NI 5734}\\\footnotesize{16-bit ADC}};
	\node [block,below=of preamp,yshift=-0.5em] (NI8133) {\textbf{NI PXIe 8133}\\\footnotesize{PC}};
	\node [block,below=of detector,yshift=-0.5em] (polaris) {\textbf{XIA DGF}\\\footnotesize{Polaris}};
    \node [block,right=of NI8133,xshift=-0.5em] (NI7966R) {\textbf{NI PXIe 7966R}\\\footnotesize{FPGA}};
	
	\path[line] (detector) -- (preamp);
	\path[line] (preamp) -- (NI5734);
	\path[line] (NI5734) -- (NI7966R);
	\path[line] (NI7966R) -- (NI8133);
	\path[draw,dashed,thick] (polaris) -- node[anchor=east]{+2500V} (detector);
	\path[draw,dashed,thick] (polaris) -- (preamp);
	
	\def\y{-1.1}
	\def\x{1.025}
	\path [draw,thick] (\x,\y) -- (0.2+\x,\y) -- (0.25+\x,\y+0.25) .. controls (0.35+\x,\y) .. (0.5+\x,\y);
	
	\def\y{-1.1}
	\def\x{3.35}
	\path [draw,thick] (\x,\y) -- (0.2+\x,\y) -- (0.25+\x,\y+0.5) .. controls (0.35+\x,\y) .. (0.5+\x,\y);
	
	\def\y{-2.6}
	\def\x{5.25}
	\path [draw,thick,dotted] (\x,\y) -- (0.2+\x,\y) -- (0.25+\x,\y+0.5) .. controls (0.35+\x,\y) .. (0.5+\x,\y);
	
	\def\y{-3.25}
	\def\x{3.45}
    \path [draw,red] (\x,\y+0.35) -- (0.6+\x,\y+0.35);
	\path [draw,thick,dotted] (\x,\y) -- (0.1+\x,\y) .. controls (0.2+\x,\y) and (0.25+\x,\y+0.5) .. (0.3+\x,\y+0.5) ..controls (0.35+\x,\y+0.5) and (0.4+\x,\y) .. (0.5+\x,\y) -- (0.6+\x,\y);
\end{tikzpicture}

%
%
%
%
		%